\documentclass{aa}

\usepackage{graphicx}
\usepackage{epsfig}

\begin{document}

   \title{The cylindrical jet base of M\,87 within 100$\mu\rm{as}$ of the central engine}

   \author{B. Punsly}

   \institute{1415 Granvia Altamira, Palos Verdes Estates CA, USA
90274: ICRANet, Piazza della Repubblica 10 Pescara 65100, Italy and
ICRA, Physics Department, University La Sapienza, Roma,
Italy\\
\email{brian.punsly@cox.net}\\ }

   \date{Received September 22, 2023; }
\titlerunning{M\,87 Jet Within 100$\mu\rm{as}$ of the Central Engine}

\abstract {A recent article on high-resolution 86 GHz observations with the Global Millimeter VLBI Array, the phased Atacama Large Millimeter/submillimeter Array, and the Greenland Telescope describes the detection of a limb-brightened cylindrical jet, $25 \mu\rm{as}< z< 100 \mu\rm{as}$, where $z$ is the axial displacement from the supermassive black hole in the sky plane. It was shown to be much wider and much more collimated than 2D simulations of electromagnetic (Blandford-Znajek) jets from the event horizon predicted. This was an unanticipated discovery. The claimed detection of a jet connected to the accretion flow provides a direct observational constraint on the geometry and physics of the jet launching region for the first time in any black hole jetted system. This landmark detection warrants further analysis. This Letter focuses on the most rudimentary properties, the shape and size of the source of the detected jet emission, the determination of which is not trivial due to line-of-sight effects. Simple thick-walled cylindrical shell models for the source were analyzed to constrain the thickness of the jet wall. The analysis indicates a tubular jet source with a radius $R\approx 144 \mu\rm{as}\approx 38M$ and that the tubular jet walls have a width $W \approx 36\mu\rm{as} \approx 9.5 M$, where $M$ is the geometrized mass of the black hole (a volume comparable to that of the interior cavity). The observed cylindrical jet connects continuously to the highly limb-brightened jet (previously described as a thick-walled tubular jet) that extends to $z> 0.65$ mas, and the two are likely in fact the same outflow (i.e., from the same central engine).}


   \maketitle

\section{Introduction}
Due to its proximity and brightness, the jet in the galaxy M\,87 (located at a distance of $\approx 16.8$ Mpc) offers the best opportunity to observe the jet launching mechanism associated with supermassive black holes. High-frequency very long baseline interferometry (VLBI) can, in principle, resolve this region. However, this is not a trivial task since the spectrum of the inner jet appears to be steep: lower frequencies are brighter \citep{had16}. Furthermore, the favorable geometry that provides the limb-brightening responsible for the brightness of the inner jet is likely lost if the jet tapers rapidly toward the central engine in the launch region. Not surprisingly, the 230 GHz VLBI observations of the
Event Horizon Telescope (EHT) have not yet detected a jet, only an annulus of emission \citep{eht19}. The annulus is also very bright, even at 86 GHz, and provides a dynamic range challenge for observations if the inner jet is not limb-brightened \citep{lu23}. However, global VLBI at 86 GHz with baselines comparable to the next-generation EHT can provide an excellent compromise between jet brightness and resolution \citep{joh23}. This was achieved in the north-south direction in a 2018 observation with the Global Millimeter VLBI Array, the phased Atacama Large Millimeter/submillimeter Array, and the Greenland Telescope \citep{lu23}. Previously, there had been little observational information on the jet configuration just above the EHT annulus and therefore little to restrict the jet produced in general relativistic magnetohydrodynamic (GRMHD) simulations. These new observations provide the strongest physical constraints to date on the jet launching mechanism. The observing team discovered a wide cylindrical shell of emission ($\sim 290 \mu\rm{as}$ in diameter) in a region $25\mu\rm{as}< z< 100\mu\rm{as}$, above the mid-line of the $42\mu\rm{as}$ diameter EHT annulus and the 86 GHz nucleus (see Fig. 1 for more clarity on the notion of a mid-line), where $z$ is the axial distance along the jet on the sky plane \citep{lu23}. This is in contrast to the de-projected (intrinsic) distance, $z'$. For the line of sight (LOS) to the jet of $18^{\circ}$ that is assumed throughout and $3.8\mu\rm{as} =M,$ where M is the geometrized mass, $z'(M)=0.838z(\mu\rm{as})$ \citep{eht19}.
\par The detection of a $\sim 76 M$ diameter, bright, cylindrical shell of jet emission, just $z'\approx 20M$ above the mid-line of the putative accretion flow associated with the 11M diameter EHT annulus, is a landmark discovery in the field of jet launching from astrophysical black holes \citep{eht19}. Since it appears to connect continuously to the ridges of the limb-brightened jet seen on larger scales, it is apparently the first direct glimpse of the physics of jet launching in a black hole accretion system. As such, its existence is worthy of a comprehensive analysis independent of theoretical biases. This study provides a modest first step in this quest. A fundamental physical aspect of this structure is the intrinsic width, $W$, of the walls of the tubular source that creates the emission of the cylindrical jet base. This is not trivially related to simple data reduction, such as the half width at half maximum (HWHM) of the emission on the outer side of the peak of the surface brightness along the ridge of the cylindrical walls. There are complicated projection effects for a nearly polar (close to the jet axis) LOS. A study of strong limb brightening farther downstream, $350\mu\rm{as}< z< 650\mu\rm{as}$, showed that the shape of the observed peaks can be used to discriminate between various simple axisymmetric geometrical models: cones, cylinders, and combinations thereof \citep{pun23}. The present study attempts to use these same methods to estimate $W$ and $R$ (the radius to the outer edge of the tubular jet wall) of the newly discovered cylindrical jet base.

\begin{figure*}
\begin{center}
\includegraphics[width= 0.75\textwidth]{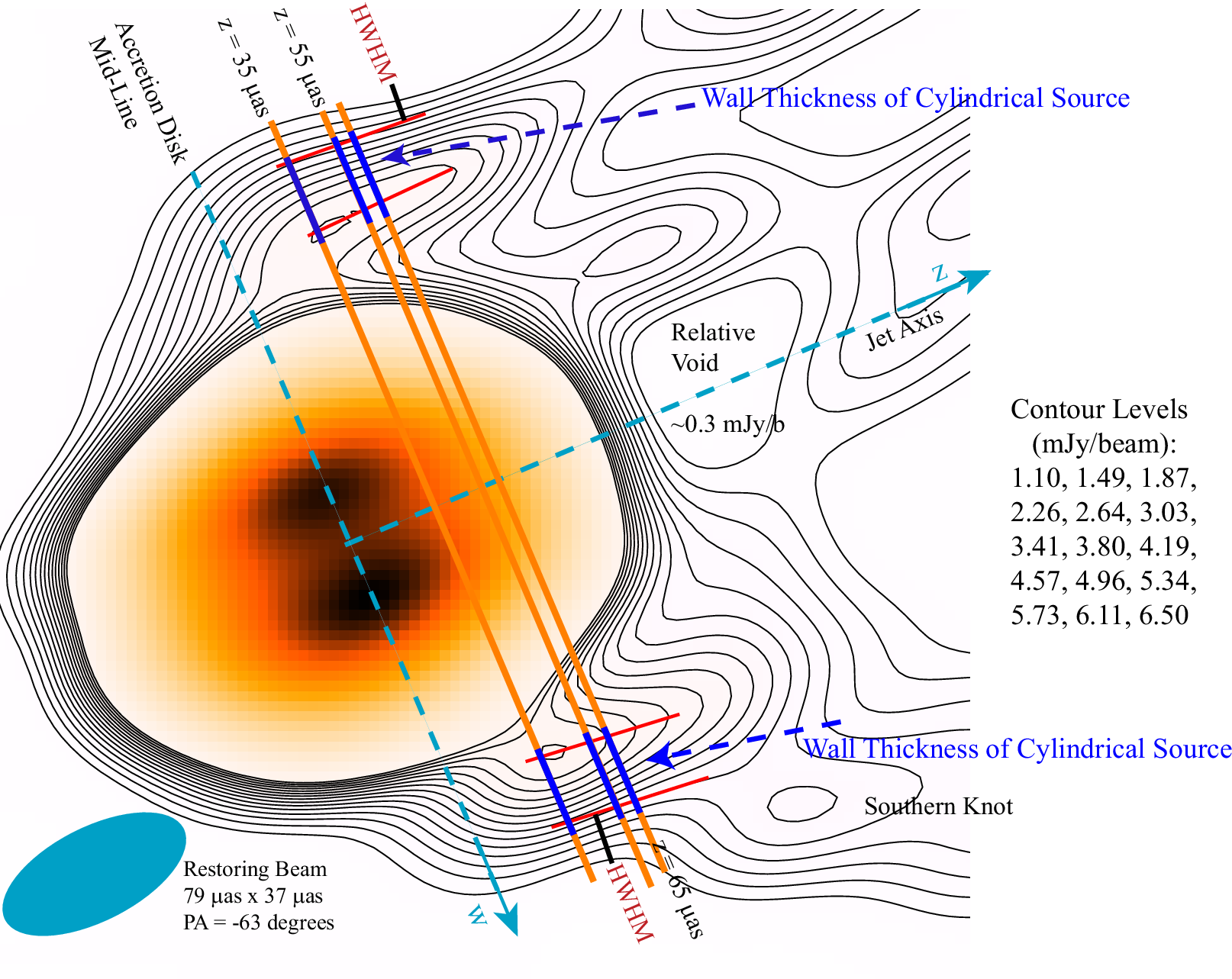}
\caption{\footnotesize{Reformatted image from the observations described in \cite{lu23}. This image highlights the cylindrical jet base at the expense of saturating the nucleus. The sky plane is the w-z plane.}}
\end{center}
\end{figure*}
\par This Letter is organized as follows. The next section shows the image from \citet{lu23}, reformatted so as to highlight the inner jet. Essential features are discussed that will guide the subsequent analysis. Section 3 reviews the key elements of the axisymmetric models with a nearly polar LOS that were discerned in \cite{pun23}. This review motivated choices regarding the features of the simple cylindrical source models of the newly discovered cylindrical jet, which are described in Sect. The implications of these results are discussed in Sect. 5.
\section{A close-up view of the cylindrical jet base}
Figure 1 is made from the same image FITS file as Fig. 1a from \cite{lu23}, restored with uniform weighting. The image was reformatted to highlight the details of the cylindrical jet. There are four significant changes:
(1) the field of view is a close-up of the jet base; (2) a linear scale is used instead of a log scale in order to track the gradients in the limb-brightened ridges more accurately; (3) the nucleus (which is not the focus of this study) is allowed to saturate at 6.5 mJy/b as opposed to 128 mJy/b, and (4) the lowest contour is 1.1 mJy/b instead of 0.5 mJy/b.

The image in \citet{lu23} was obtained by averaging five CLEAN images that were independently generated. Inspection of an individual image FITS file (shared by Rusen Lu) shows numerous contours at -0.7 mJy/b to -0.85 mJy/b $\sim 0.6$ mas from the nucleus. This motivated the conservative choice of $\gtrsim 1$ mJy/b for the lowest (reliable) contour level in Fig. 1.
\par Various details are highlighted in the image. First is the mid-line of the putative accretion disk. Near the source, projection effects can make the base of a concave shell (i.e., a parabolic) on the approaching jet side appear below the mid-line, and it is not obvious how to interpret the z coordinate (see, e.g., Fig. 10 of \citealt{cha19}). Unambiguous coordinates are needed in order to be quantitative, and the mid-line is the choice for the transverse axis. The transverse coordinate is labeled "w" and the axial coordinate "z." Second, this study explores the three cross-cuts indicated by the orange lines. Third, the red lines are linear approximations of the peak intensity of the ridges (inner line) and the HWHM of the ridge (outer line) in the range $30 \mu\rm{as}< z< 95 \mu\rm{as}$. The outer red line is defined similarly as that used in Fig. 3 of \cite{lu23} and agrees closely with their result. Their plot continues to smaller z from considerations of a super-resolved image. The smallest z for which the HWHM can be defined without super-resolution is $30 \mu\rm{as}$. The north (south) ridge has an intensity in the range $457\pm39$ mJy/b ($543\pm68$ mJy/b) for $20 \mu\rm{as}< z< 100 \mu\rm{as}$. The north (south) linear fit to the intensity maxima, $30 \mu\rm{as}< z< 95 \mu\rm{as}$, is at PA= $-65^{\circ}$ (PA= $-73^{\circ}$). Similarly, the north (south) linear fit to the HWHM, $30 \mu\rm{as}< z< 95 \mu\rm{as}$, is at PA= $-70.5^{\circ}$ (PA= $-72^{\circ}$). The jet axis was chosen as PA= $-67^{\circ}$, the same as the jet direction estimated previously on sub-mas scales \citep{had16}. Based on the above, a uniform cylindrical shell, $20 \mu\rm{as}< z< 100 \mu\rm{as}$, is suggested as the simplest (least fine-tuned) model to try.
\par The meaning of the distance from the ridge to the HWHM point has no obvious connection to the physical source in a nearly polar LOS geometry. The primary goal here is to determine the physical source dimensions that generate the transverse profiles of the ridges. This is estimated in Sect. 4 for the three cross-cuts. The estimated $W$ are indicated by the three dark purple transverse segments in Fig. 1.

\section{Review of axisymmetric source models}
Figure 2 highlights the essential elements of cross-sections produced by a nearly polar LOS to an axisymmetric luminous volume. It is a modified version of Fig. 2 from \citet{pun23}. It is a more general geometry than the cylinder considered here; the cone near the nucleus is no longer modeled. The central intensity trough (for a limb-brightened image) is controlled by contributions from LOS exit points from the jet wall (close to the nucleus near x=0 in Fig. 2). The nature of this putative region cannot be determined from the observation since the surface brightness is too low. This is not an issue for this analysis, since more than half of the intensity cross-cuts, which are shrouded by a nuclear glow (i.e., the region near x=0 in the middle of the cross-cut), were not modeled. More importantly, for the high transverse resolution of this observation, these LOSs have no effect on the shape of the outer portion of the ridges.

\begin{figure*}
\begin{center}
\includegraphics[width= 0.6\textwidth]{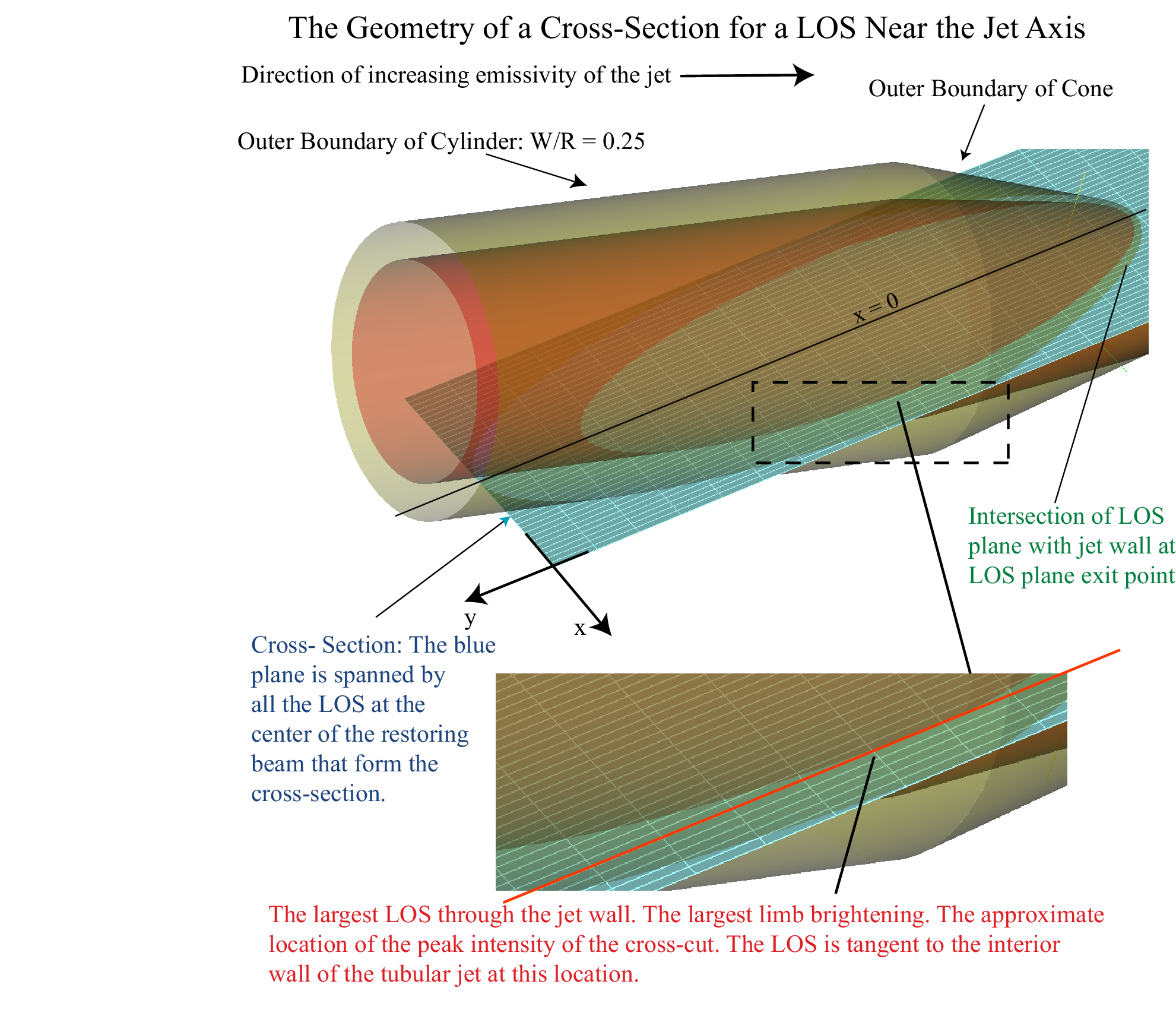}
\caption{\footnotesize{Blue (x, y) plane foliated by the set of all LOSs that constitute a cross-cut. The intersection of the plane with the tubular jet is in dark olive green. A zoomed-in view of the region of maximum limb-brightening, delimited by the dashed black rectangle, is shown at the bottom. This region has the longest LOS through the tubular jet wall (red line), as indicated by the white grid overlaid on the blue LOS plane.}}
\end{center}
\end{figure*}
\par Another important feature is the cutout of the region with strong limb brightening. The longest path length occurs when the LOS grazes the inner edge of the tubular jet. The peak of the cross-cut is always near the location on the sky plane for which the LOS through the center of the restoring beam is along this trajectory. As a corollary to this, the LOS path lengths decrease as the beam center moves toward the outer edge of the jet. When the transverse resolution is comparable to or smaller than $W$, and the beam center is at the outer edge of the jet wall, the flux density will be $<0.5$ of the peak intensity due to the shorter path lengths through the jet wall. The outer edge of the jet will necessarily be farther from the peak than the HWHM (as in Fig. 1).

\begin{figure*}
\begin{center}
\includegraphics[width= 0.35\textwidth]{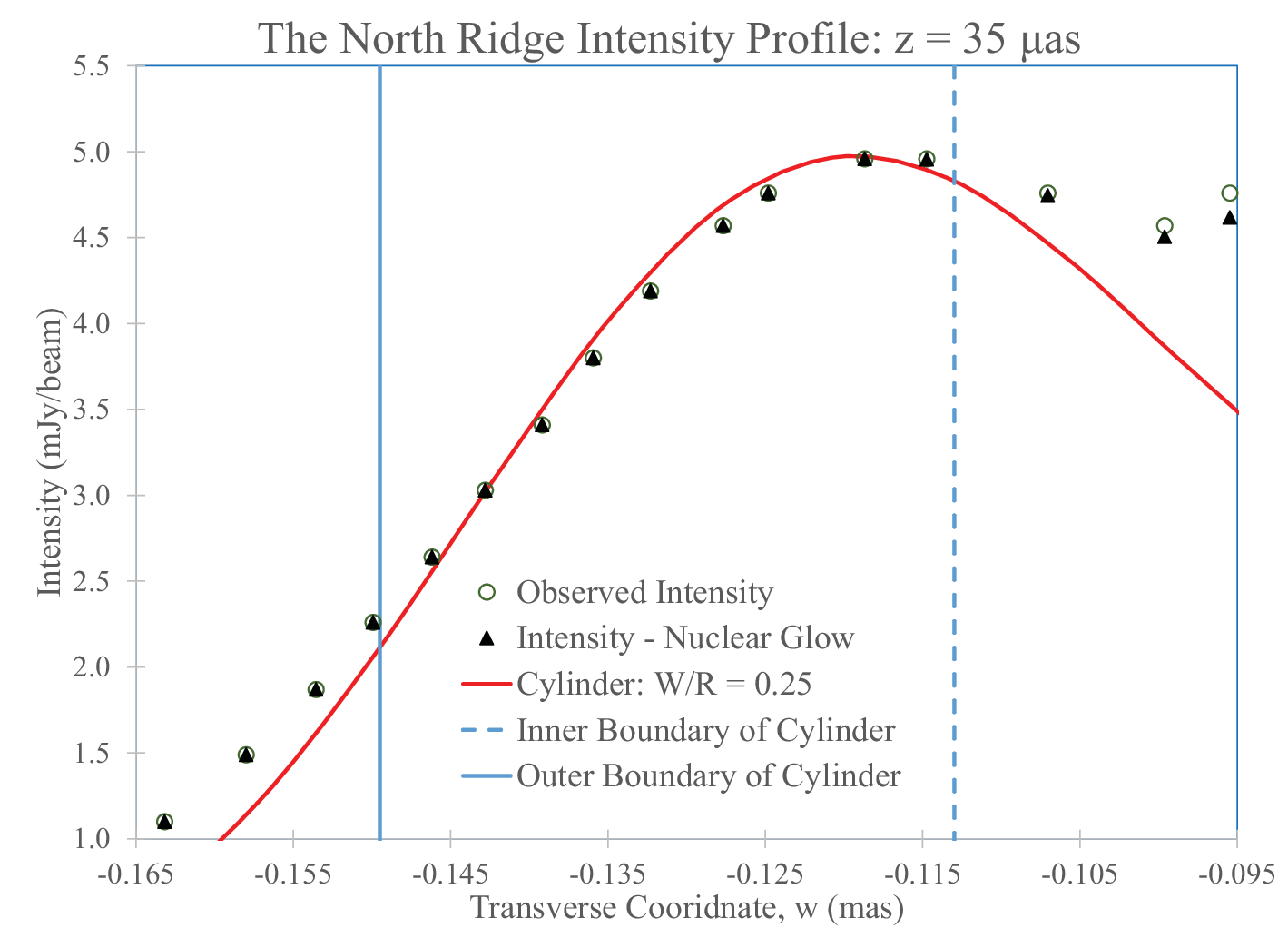}
\includegraphics[width= 0.35\textwidth]{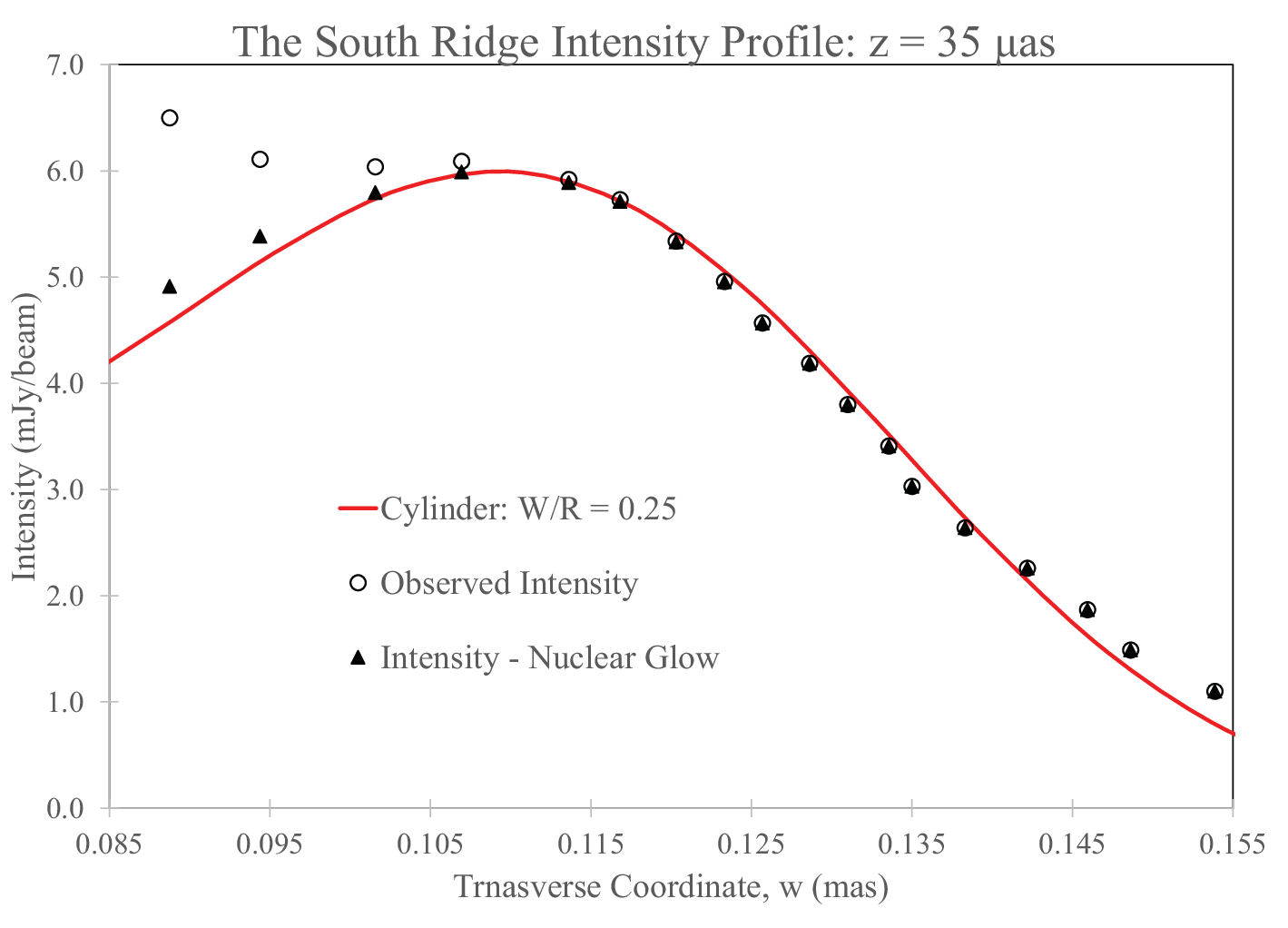}
\includegraphics[width= 0.35\textwidth]{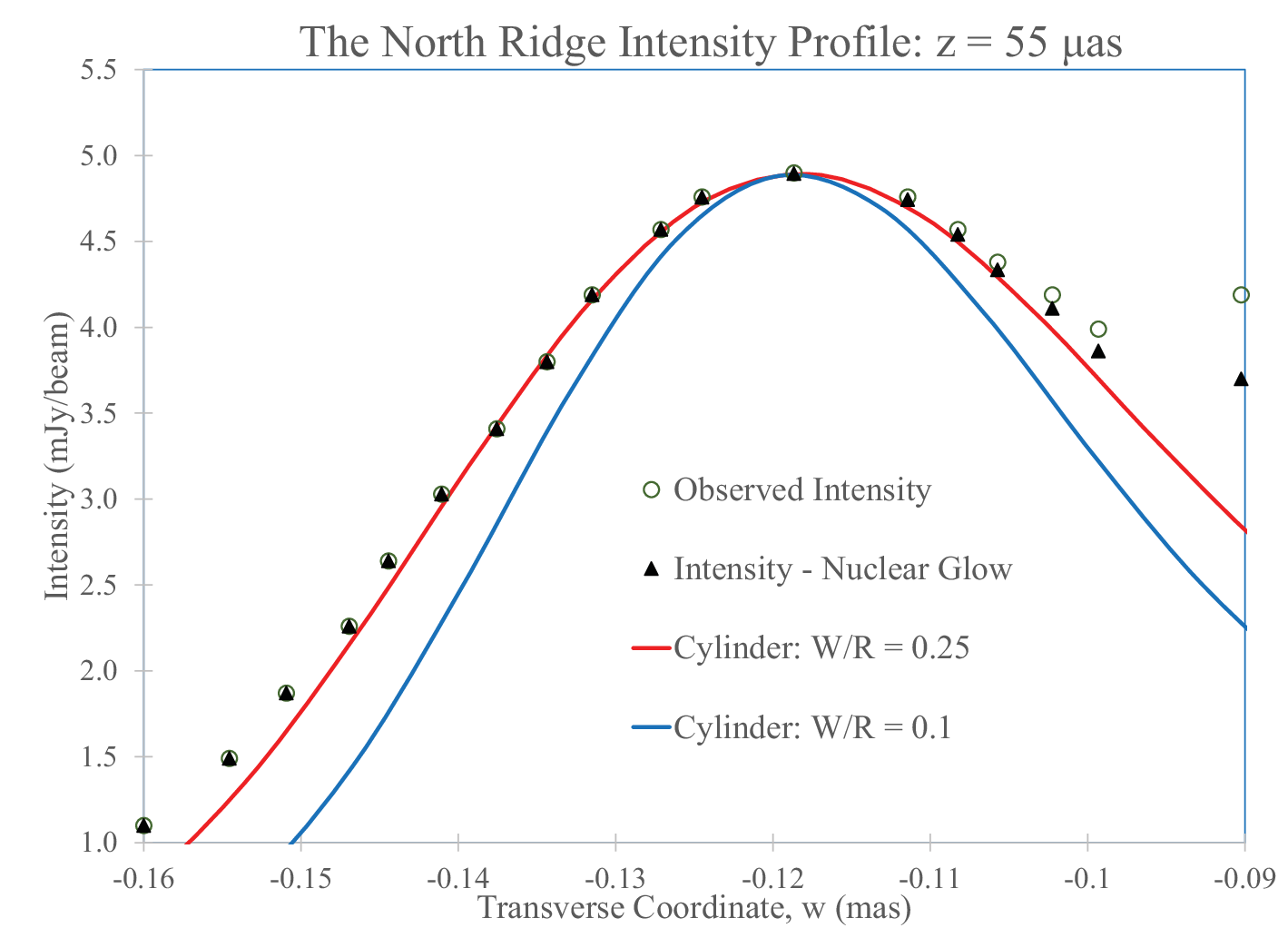}
\includegraphics[width= 0.35\textwidth]{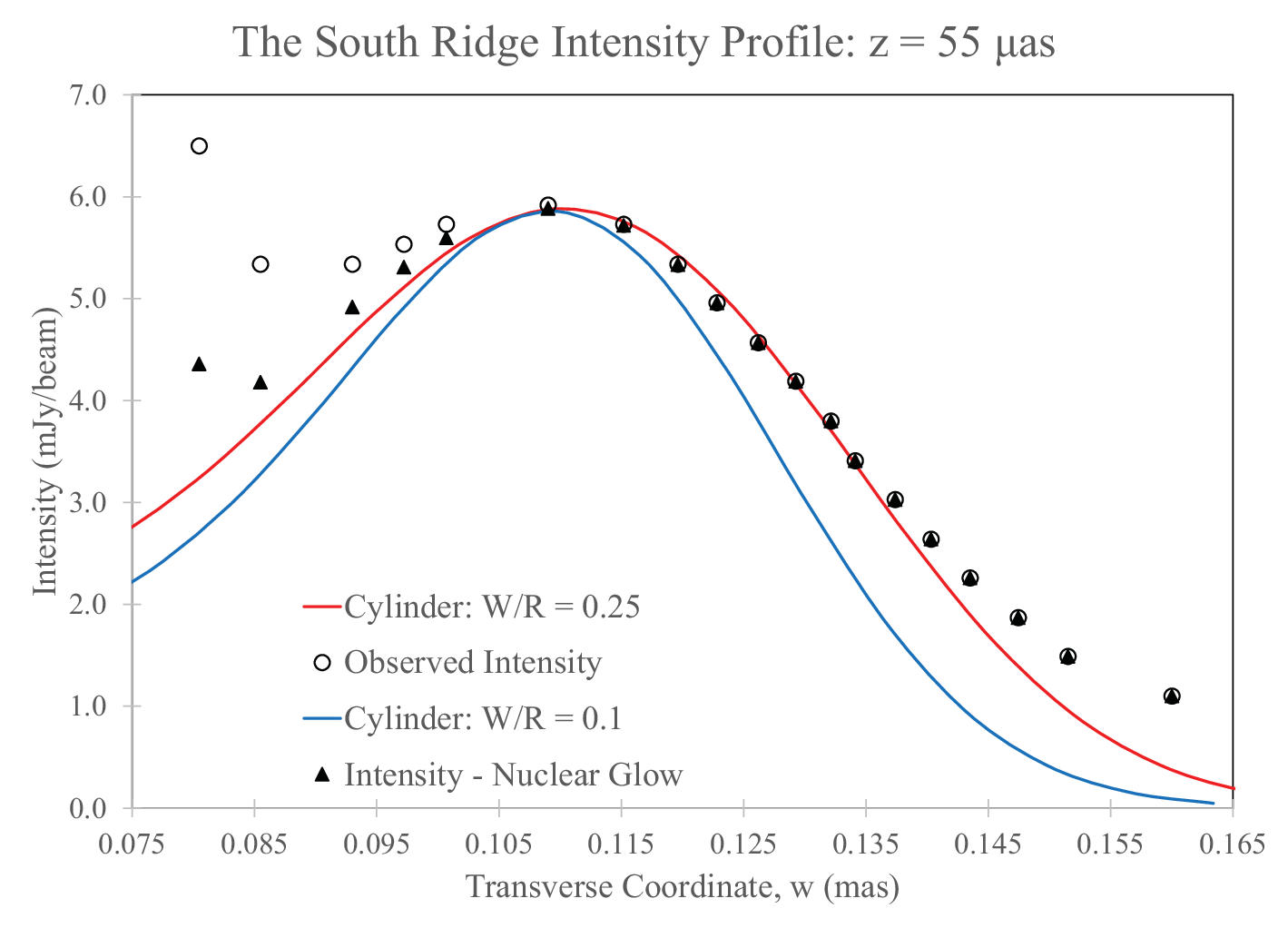}
\includegraphics[width= 0.35\textwidth]{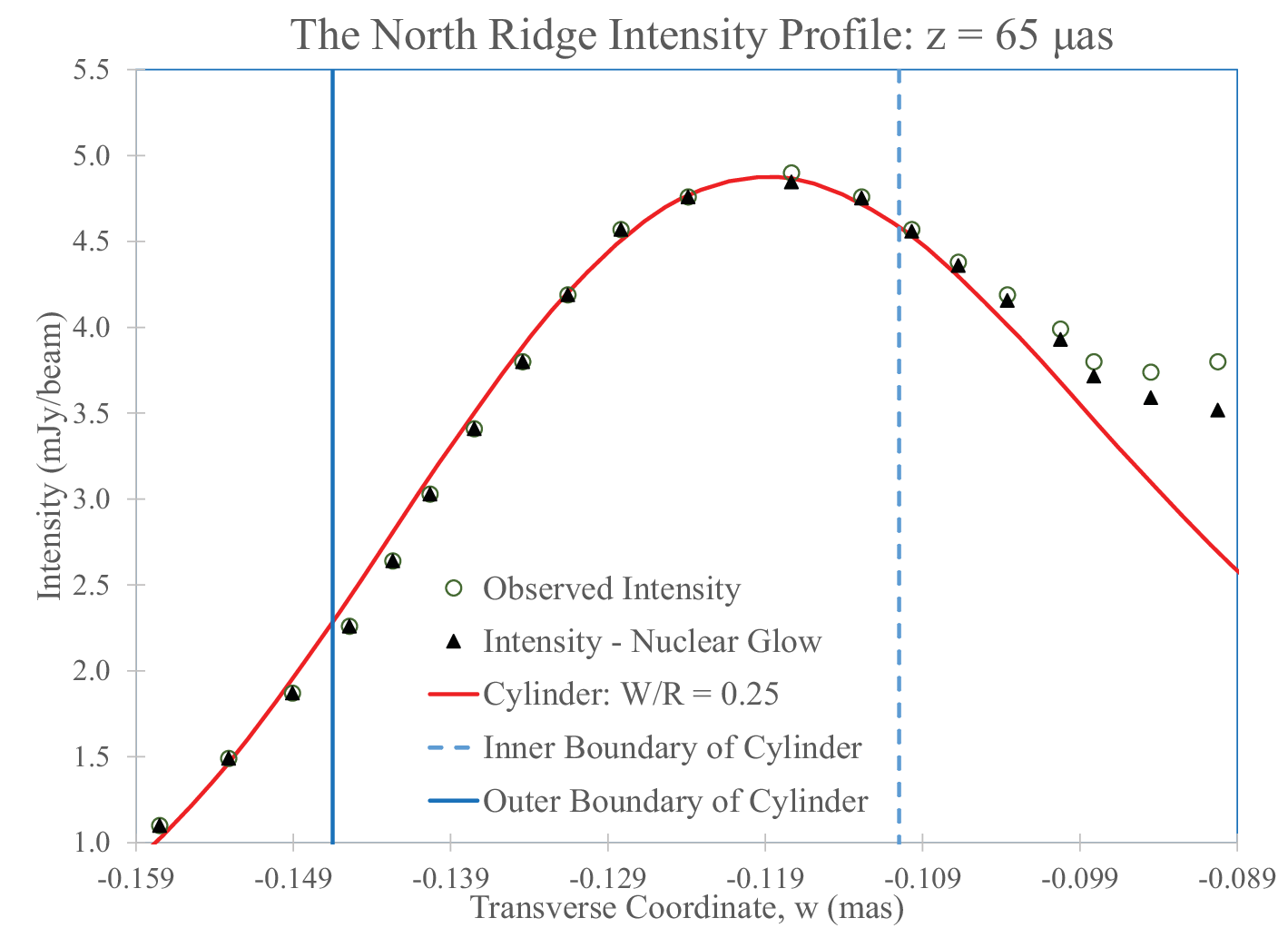}
\includegraphics[width= 0.35\textwidth]{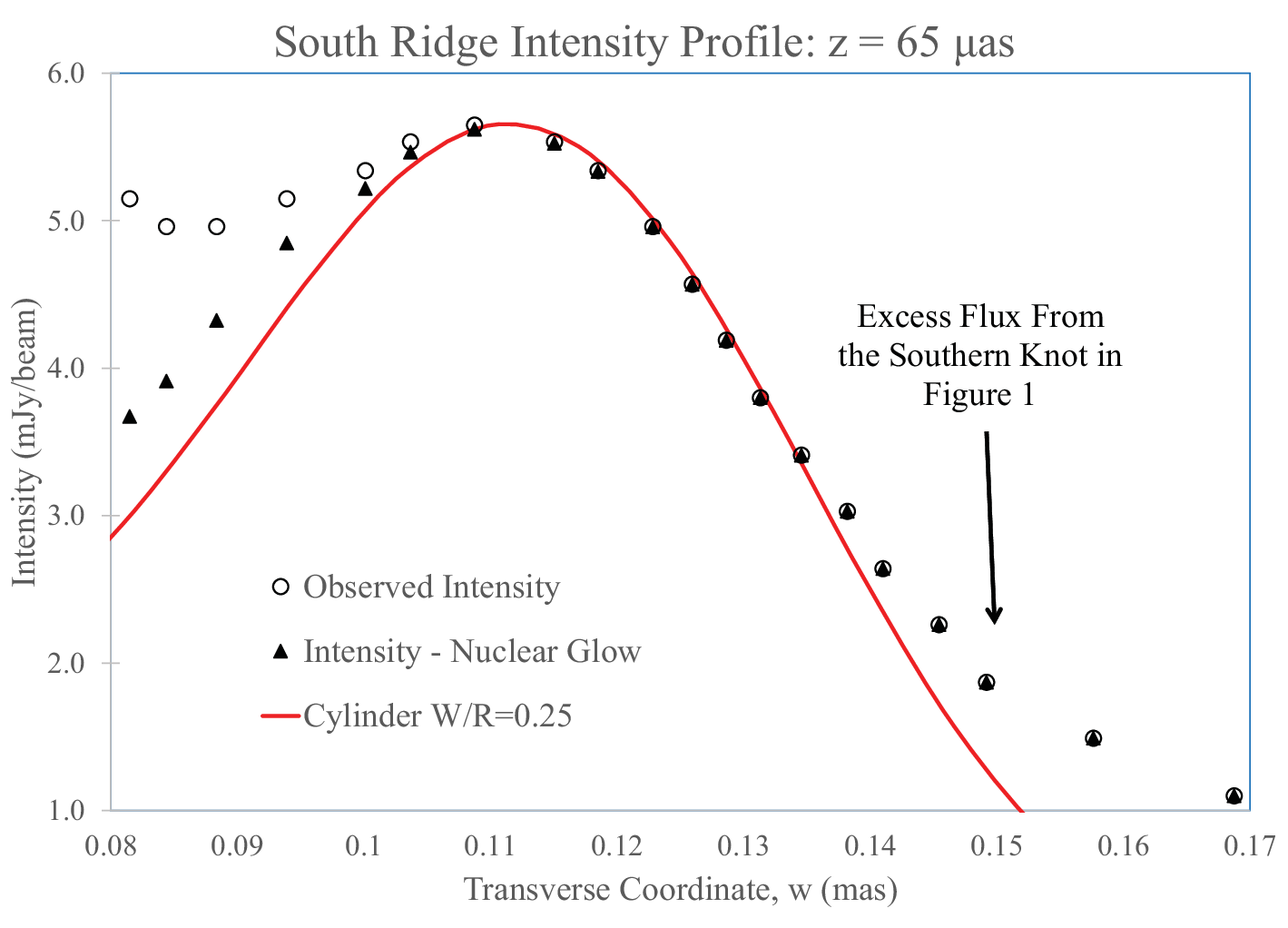}
\caption{\footnotesize{Cross-sections of a uniform cylindrical shell after convolution with the restoring beam compared with actual cross-sections of the cylindrical jet base. The emissivity is constant in the range $20 \mu\rm{as}< z< 100 \mu\rm{as}$, then fades linearly to 25\% of this value in the range $100 \mu\rm{as}< z< 200 \mu\rm{as}$,} and is zero elsewhere. The middle frame indicates that W/R=0.1 is too narrow.}
\end{center}
\end{figure*}

\section{A simple cylindrical source for the cylindrical jet base}
In Sect. 2 it was noted that the linear fits to the HWHM and intensity maxima demarcate fairly constant separations on the sky plane, $30 \mu\rm{as}< z< 95 \mu\rm{as}$. Also, the surface brightness is constant in the north (south) ridge to within $\pm8.5\%$ ($\pm12.5\%$), $20 \mu\rm{as}< z< 100 \mu\rm{as}$. This calls for a very simple model, a uniform cylinder ($20 \mu\rm{as}< z< 100 \mu\rm{as}$), to assess the most basic geometric properties of the source of the cylindrical jet base. Axisymmetry is not perfect: the south ridge is $14\%\pm 4\%$ brighter than the north ridge, likely a consequence of mild Doppler boosting, as shown in 3D simulated emissivity distributions \citep{cha19,cru21}. Since there is a large gap that isolates the two ridges, the continuous gradient in the Doppler enhancement factor (which is unknown) was approximated from the discrete difference in the peak intensities of the ridges. Another simplification is that the major axis of the restoring beam is approximately aligned with the cylinder walls.
\begin{figure*}
\begin{center}
\includegraphics[width= 0.35\textwidth]{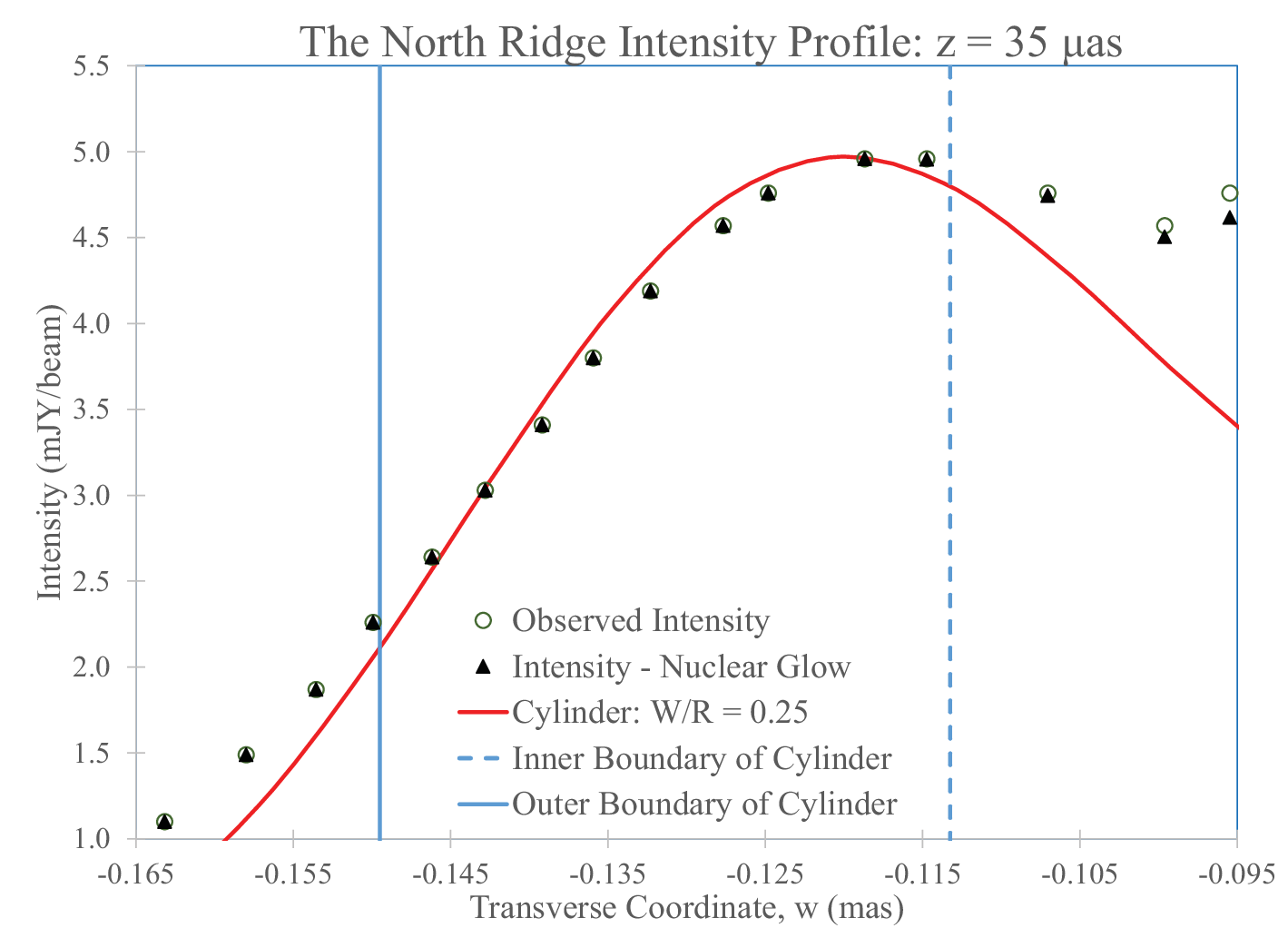}
\includegraphics[width= 0.35\textwidth]{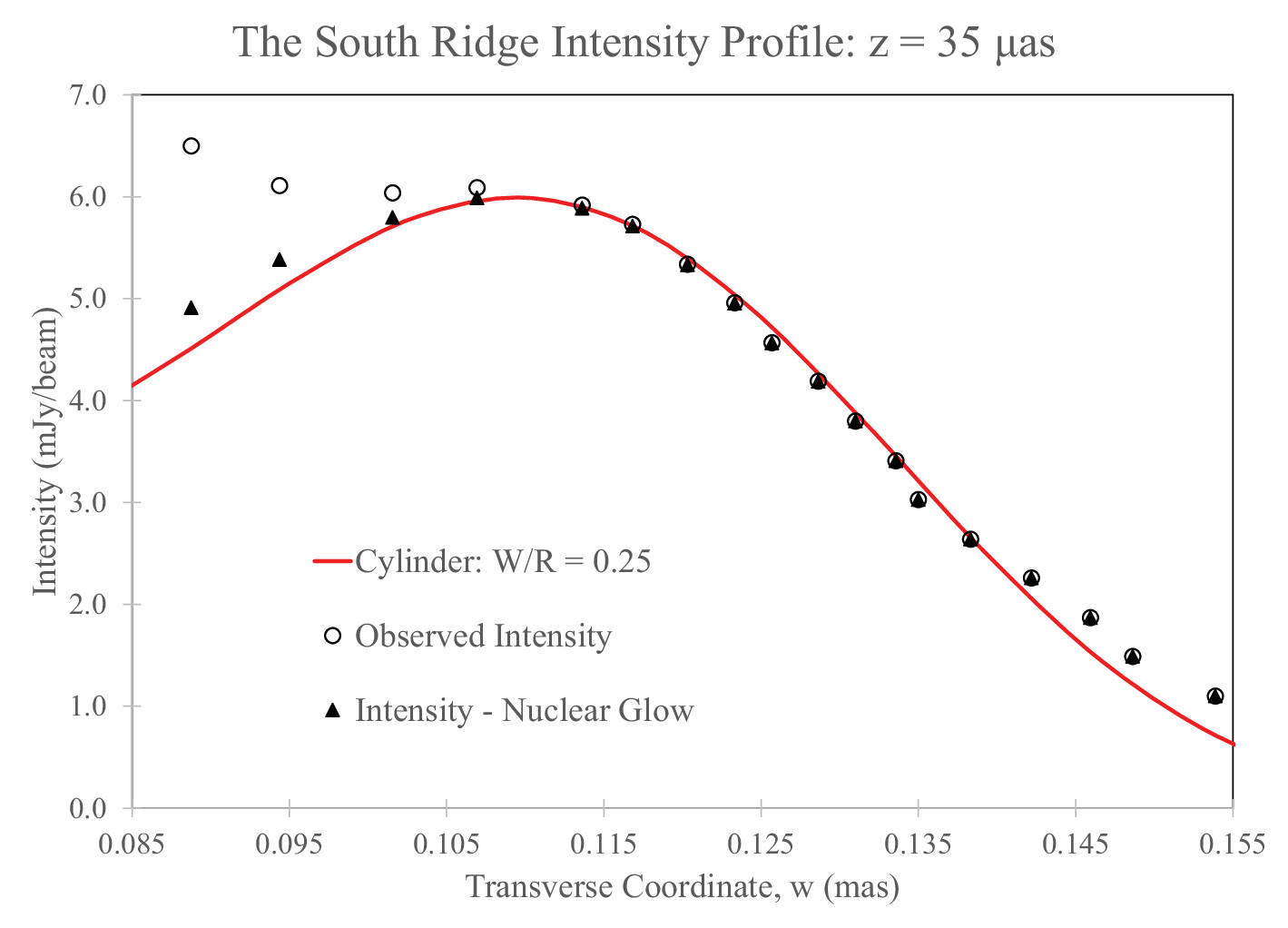}
\includegraphics[width= 0.35\textwidth]{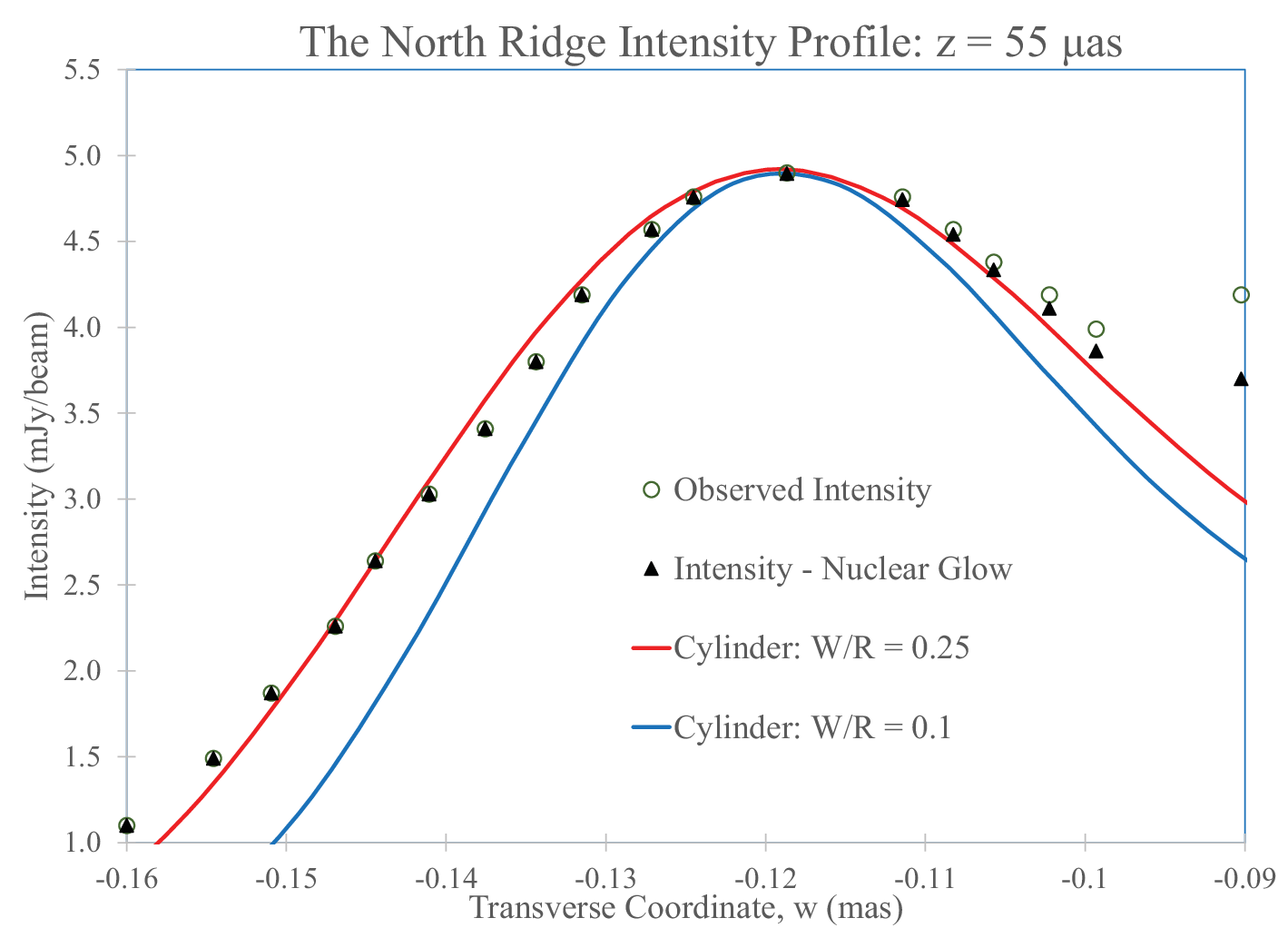}
\includegraphics[width= 0.35\textwidth]{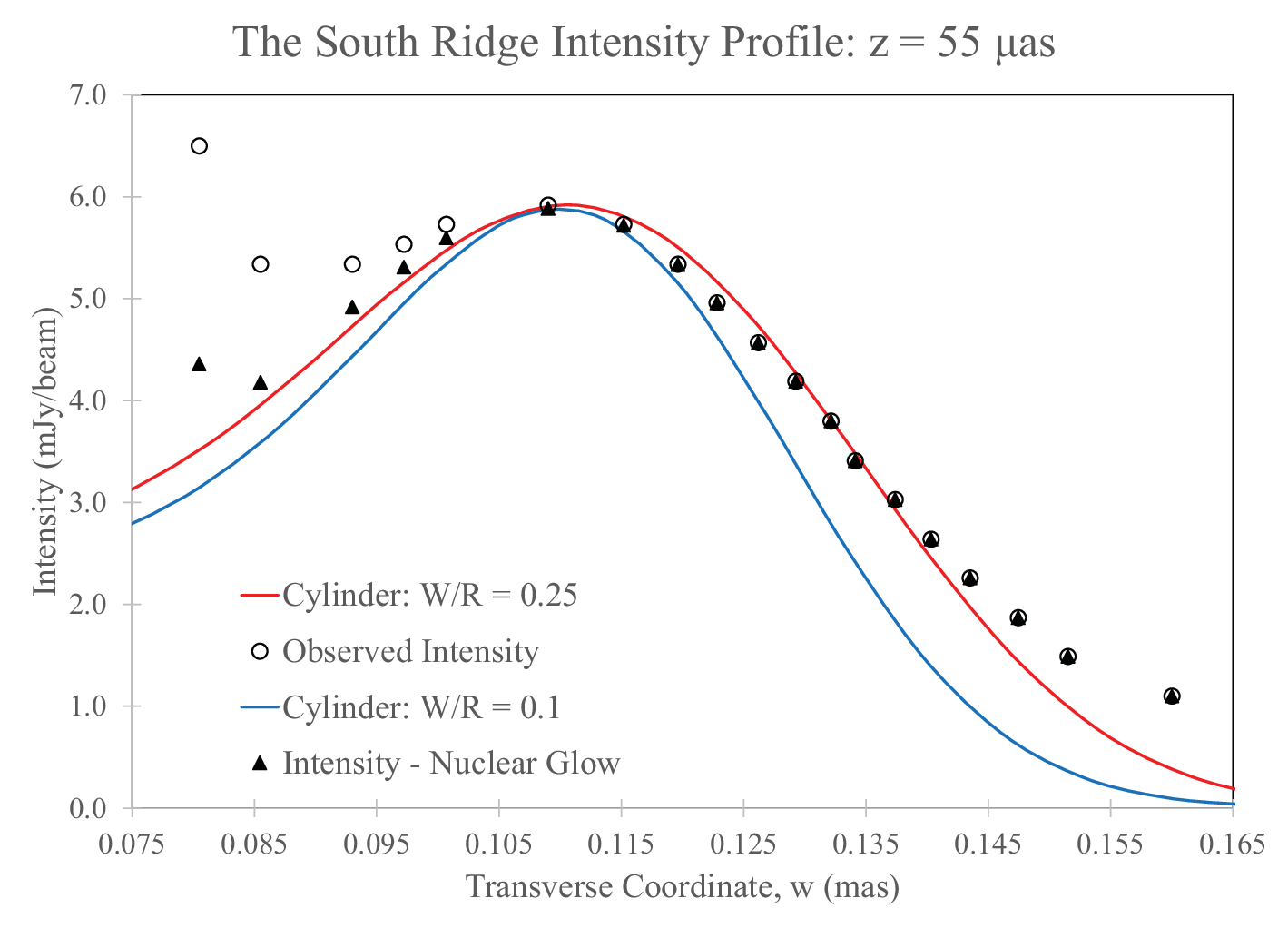}
\includegraphics[width= 0.35\textwidth]{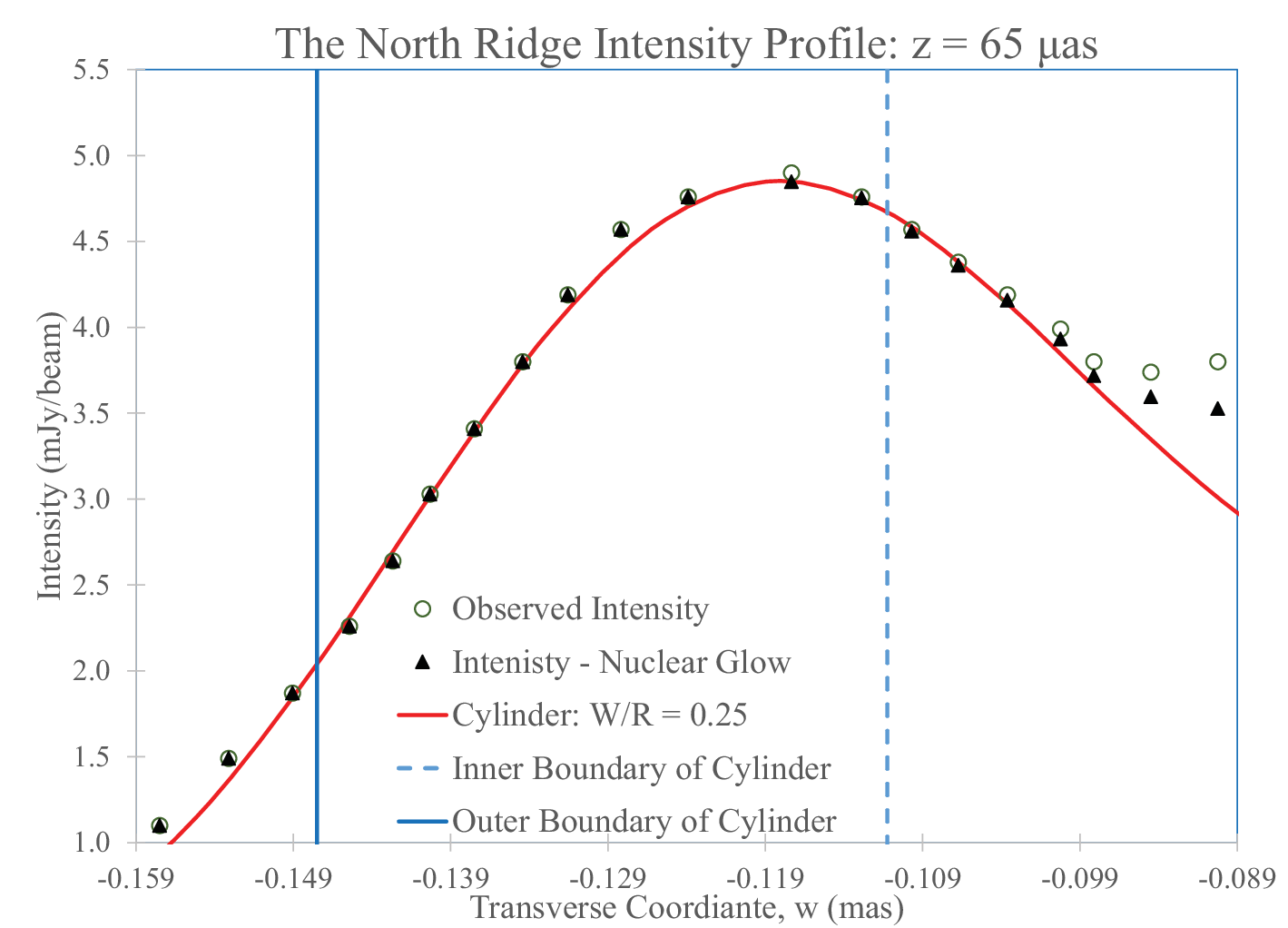}
\includegraphics[width= 0.35\textwidth]{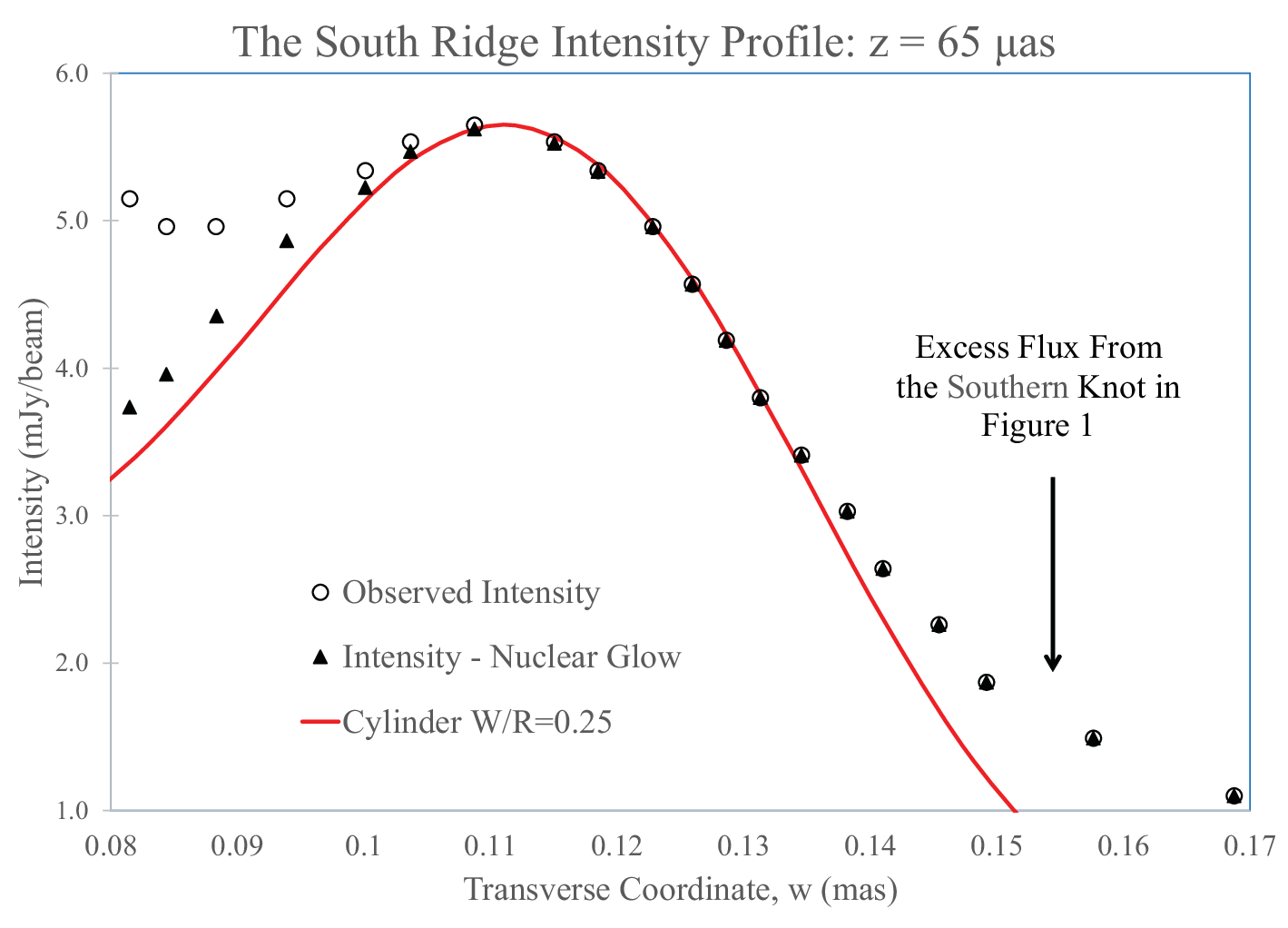}
\caption{\footnotesize{Same as Fig. 3 except that the emissivity is constant for  $15 \mu\rm{as}< z< 200 \mu\rm{as}$ and zero otherwise.}}
\end{center}
\end{figure*}
\par Figure 3 shows cross-sections at the three locations indicated in Fig. 1. These ridges are resolved from the bright nucleus (i.e., there is a minimum between the ridge peak intensity and the strong glow of the nucleus in Fig. 1). The first step was to subtract off the nuclear glow. The nucleus appears as a double component in Fig. 1. Therefore, it was fit by two Gaussian functions. The nuclear glow does not affect the peak or the outer HWHM in Fig. 3. There is a correction on the inward side of the peak. It is not very accurate because there is nuclear emission in excess of the nuclear double Gaussian, centered at $w\sim \pm 65\mu\rm{as}$, as evidenced by the super-resolved image shown in Fig. 1b of \citet{lu23}. In Figs. 3 and 4 of this Letter, this excess creates an inflection point in the nuclear-glow-corrected intensity, inward of the peak. No real faint excess from an artifact can be unambiguously distinguished from  these observations due to the adjacent bright core, so this was not pursued further. In this Letter, the nuclear subtraction is considered a crude improvement on the inward side of the peak. The top-left panel of Fig. 3 indicates that the nuclear contamination limits the cross-sectional analysis to $z\geq35 \mu\rm{as}$.

\par The cylinder is terminated at $z=20 \mu\rm{as}$ in the model used to create Fig. 3. Looking at Fig. 1, there is not much evidence of significant flux inside of $z=20 \mu\rm{as}$. Extrapolating the linear fit to the ridge maxima to $z=20 \mu\rm{as}$ yields 476\,mJy/b (554\,mJy/b) in the north (south) and half this value is achieved $45 \mu\rm{as}$ ($38 \mu\rm{as}$) away. Since the restoring beam has a HWHM of $\approx 39.5 \mu\rm{as}$ in this direction, the data are consistent with minimal or no emissivity at $z<20 \mu\rm{as}$ along this line.
\par Another issue is the emissivity where the LOS enters the tubular jet. We can see from Fig. 2 that, due to the nearly polar LOS, this is very far from where the LOS exits the jet. Since the cylindrical source is convolved with the restoring beam (0.079 mas major axis), one must consider LOSs that enter the tubular jet out to $z\sim200 \mu\rm{as}$. Extrapolating the region demarcated by the HWHM and the peak from $100 \mu\rm{as}$ to $200 \mu\rm{as,}$  Fig. 1 indicates a decrease in intensity. The peak intensity in the north (south) drops from the median value at $20 \mu\rm{as}< z< 100 \mu\rm{as}$ (found in Sect. 2) to 41\% (28\%) of this level at $z=200 \mu\rm{as}$. It would be unrealistic to cut off the cylinder at $z=100 \mu\rm{as}$. There is clearly jet emission beyond this in Fig. 1, but it is not represented by a uniform cylinder. For calculational simplicity, a very crude fit with a cylinder (same R and W as $20 \mu\rm{as}< z< 100 \mu\rm{as}$) was chosen for this region under the hypothesis that it is a far-field, lower-order contribution to the cross-sections. This hypothesis was tested by choosing two extremes in the cylinder emissivity for $100 \mu\rm{as}< z< 200 \mu\rm{as}$ to investigate if there is any significant difference. If there is not, the method and hypothesis are reasonable. Figure 3 uses an aggressive model of the emissivity tapering. The emissivity is constant in the range $20 \mu\rm{as}< z< 100 \mu\rm{as}$. For $z>100 \mu\rm{as}$ it decreases linearly to 25\% of the value at $z = 200 \mu\rm{as}$. Figure 4 depicts the opposite truncation extreme. The emissivity is constant for $15 \mu\rm{as}< z< 200 \mu\rm{as}$. Comparing Figs. 3 and 4, we see that the differences in the fits are small, except for the cross-section at $z=65 \mu\rm{as}$. The cross-section of the southern ridge is modestly affected by the truncation inside of the peak. This is the reason why $z\leq65 \mu\rm{as}$ bounds the cross-sections in this analysis. It is concluded that the truncation does not strongly affect the fits to the cross-sections and that the primary contributor is the uniform cylinder, $20 \mu\rm{as}< z< 100 \mu\rm{as}$.
\par The main result of the analysis is shown in the middle panels of Figs. 3 and 4, the $z=55 \mu\rm{as}$ cross-section. This is the best compromise between nuclear dilution and the effects of truncation. Irrespective of the truncation, $W/R=0.25$ is a much better source model than $W/R=0.1$. The cylindrical fits to the three cross-sections have $R\approx 144 \mu\rm{as}\approx 38M$ and $W\approx 36 \mu\rm{as} \approx 9.5 M$, and $W/R=0.25$ is substantially thick. The tubular wall comprises $1-(1-0.25)^{2}=0.44$ of the total volume enclosed by the jet.

\section{Discussion and conclusion}
The methods of this analysis are intended to constrain the rudimentary properties of the source of the cylindrical jet base assuming a minimal number of fine-tuned degrees of freedom. A simple uniform cylindrical shell is motivated by the relatively uniform properties of the observed cylinder (as discussed in Sect. 2). Figures 3 and 4 and the related analysis demonstrate that the observed cylindrical jet base of M\,87, $25 \mu\rm{as}< z< 100 \mu\rm{as}$, can arise from a cylindrical shell of emission if the shell is relatively thick, namely $W/R\approx 0.25,$ and not thin, $W/R\approx 0.1$. The fits to the cross-sections in the range $35 \mu\rm{as}< z< 65 \mu\rm{as}$ (a de-projected distance from the mid-line of the nucleus of $29M< z'< 54M$), shown in Figs. 3 and 4, demonstrate that this deduction is independent of the method of truncating the cylindrical volume.
\par Figure 1 and the wide field of view in Fig. A.1 indicate that the north ridge of the cylindrical jet bends northward at $z\gtrsim 100\mu\rm{as}$ toward the next knot $> 150\mu\rm{as}$ downstream. This bending is too far from the next knot to be the result of a blurring of the $\approx 5.5\rm{mJy/b}$ peak intensity knot by the restoring beam. The south ridge is contiguous to the next knot downstream (the ``southern" knot in Figs. 1 and A.1). These knots mark the start of a limb-brightened diverging region of the jet. The diverging jet appears to be an extension of the cylindrical jet. This expanding jet was shown in Fig. 1 of \citet{lu23} to reach $z>0.8\rm{mas}$. This interpretation is consistent with the finding that the source of the limb-brightened jet from $0.35 \rm{mas}< z< 0.65 \rm{mas}$ is a tubular jet with $W/R\approx 0.35$ \citep{pun23}. There is nothing in the observations to indicate anything except that the tubular jet is well formed and thick-walled by $z\sim25 \mu\rm{as}$ and propagates continuously as such as it flows to larger $z$. To provide theoretical context, it was demonstrated in Fig. 3 of \citet{lu23} that the cylindrical jet base is much wider and much more collimated than 2D-simulated Blandford-Znajek jets from rapidly rotating black holes (those that produce strong jets) for $z<65 \mu\rm{as}$ \citep{bla77}. This appears to be the case as well in 3D GRMHD based on the published emissivity distributions produced from simulations of rapidly rotating black holes \citep{cha19,cru21,fro21}. Thus, the observations indicate:
\begin{enumerate}
\item The notion that the limb-brightened portion of the jet (the sheath) is located on the outer streamlines of the Blandford-Znajek jet is not supported by the observations because the base of the continuous limb-brightened jet is much wider than these streamlines \citep{nak18}.
\item The bright limbs are not the boundary layer between the Blandford-Znajek jet and an enveloping wind. There is a large gap between the outer streamlines of the Blandford-Znajek jet and the cylindrical jet base for $z<65 \mu\rm{as}$ \citep{lu23}. There is no contact between the two and hence no boundary layer. Yet, the jet already has a thick wall at $z< 30\mu\rm{as}$ based on the HWHM at $z= 30\mu\rm{as}$ and the cross-sections of the uniform cylinder model.
\item The bright limbs and thick jet walls, $0.35 \rm{mas}< z< 0.65 \rm{mas}$, are not created by a Kelvin-Helmholtz instability between the relativistic jet and the surrounding galactic medium or wind \citep{bic95}. The thick walls already exist at $z < 30 \mu\rm{as}$. They are created at the source.
\item The observed cylindrical jet connects continuously to the limb-brightened jet. These are two regions of the same jet originating from the same central region.
\item There is no counter-jet detected. From standard Doppler beaming arguments, the cylindrical jet is at least mildly relativistic if one assumes bilateral symmetry \citep{lin85}.
\end{enumerate}
The analysis in this Letter is very basic and simple, yet there seems to be many important implications. These results should motivate future numerical model building.

\begin{acknowledgements}
I would like to thank Rusen Lu and Thomas Krichbaum for sharing their image FITS file and discussing their results with me. This manuscript benefitted from the improvements suggested by the supportive referee.
\end{acknowledgements}

\begin{appendix}
\section{Additional figure}
Figure A.1 supports the claim that the cylindrical model that is described in the main text represents the base of the larger-scale ridges of the limb-brightened jet.
\begin{figure*}
\begin{center}
\includegraphics[width= 0.75\textwidth]{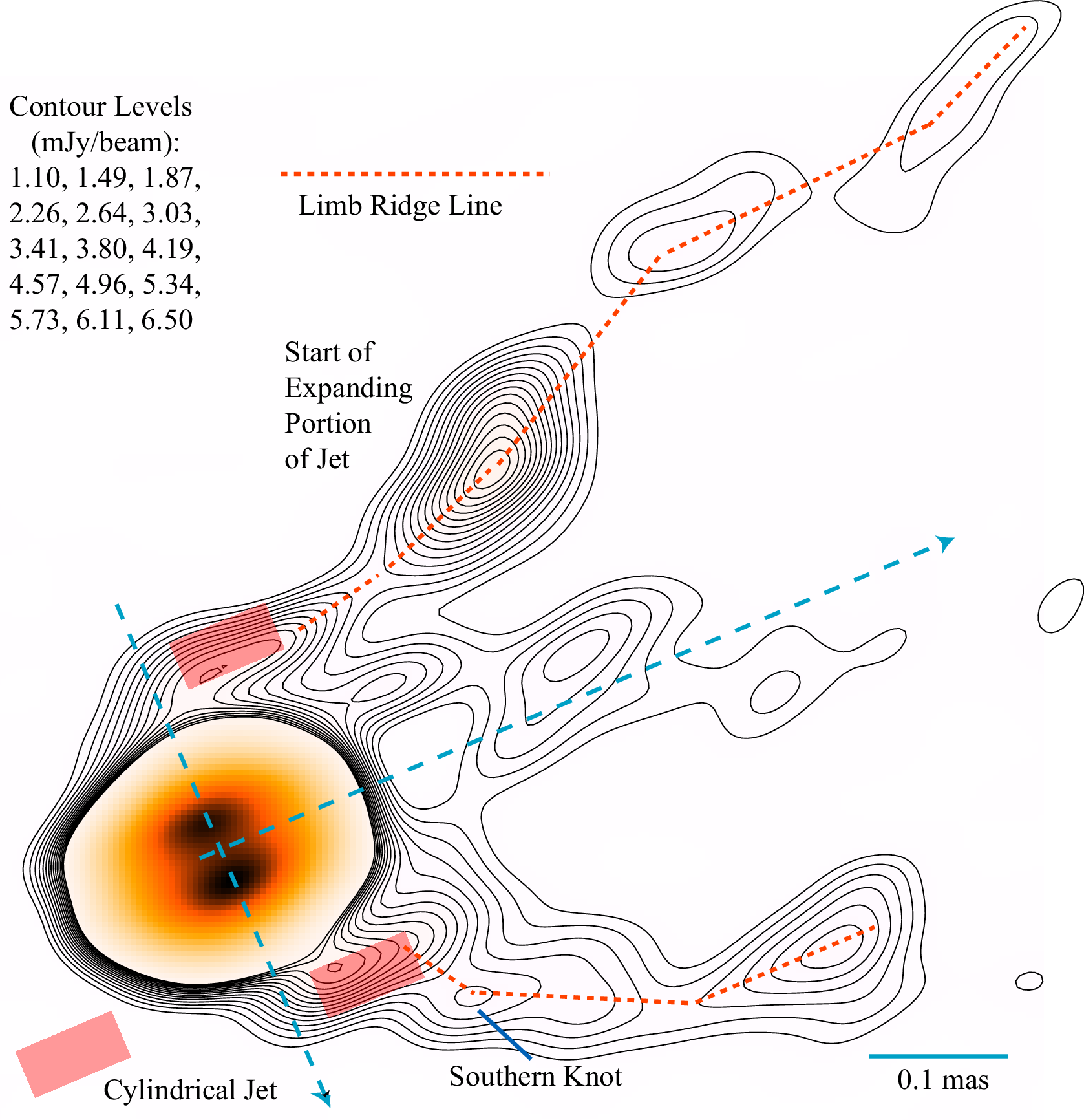}
\caption{\footnotesize{Profile of the jet limbs from $z= 25 \mu\rm{as}$ to $z> 500 \mu\rm{as}$. The contours are the same as in Fig. 1 but the field of view is wider. The dashed red limb ridge line is a linear piece-wise approximation of the peak intensity. The pink rectangles represent the cylindrical model described in the text.}}
\end{center}
\end{figure*}
\end{appendix}
\end{document}